\documentclass[%
reprint,
amsmath,amssymb,
aps,
]{revtex4-2}
\usepackage{amsmath}
\usepackage{multirow}
\usepackage{comment}
\usepackage[colorlinks=true, linkcolor=blue, citecolor=blue]{hyperref}
\usepackage{graphicx}
\usepackage{dcolumn}
\usepackage{bm}

\begin{document}
	
	\preprint{APS/123-QED}
	
	\title{Spectral Geometry and Dispersion-Constrained Projection of Diffusive Fields}
	
	\author{Pengfei Zhu}
	
	\thanks{pengfei.zhu@bam.de}
	\affiliation{%
		Bundesanstalt für Materialforschung and -prüfung (BAM), 12205 Berlin, Germany
	}%
	
	\author{Julien Lecompagnon}
	\affiliation{%
		Bundesanstalt für Materialforschung and -prüfung (BAM), 12205 Berlin, Germany
	}%
	
	\author{Philipp Daniel Hirsch}
	\affiliation{%
		Bundesanstalt für Materialforschung and -prüfung (BAM), 12205 Berlin, Germany
	}%
	
	\author{Mathias Ziegler}
	\affiliation{%
		Bundesanstalt für Materialforschung and -prüfung (BAM), 12205 Berlin, Germany
	}%
	
	\date{\today}

\begin{abstract}
	Diffusive fields obey operator-imposed relations between spatial structure and temporal decay, yet conventional spectral filtering selects components primarily according to frequency or wavenumber magnitude. Here we show that the diffusion operator defines a spectral geometry in the joint space of spatial wavenumber and modal decay rate, where physically admissible modes occupy the manifold $\eta=\alpha|\mathbf{k}|^2$. This geometry separates spectral scale from physical consistency: high-wavenumber modes can remain diffusion-consistent, whereas lower-wavenumber modes can violate the governing dynamics. We exploit this distinction by introducing an operator residual and a finite-width soft projection that selects spectral components according to their distance from the diffusion manifold rather than their spectral magnitude. Numerical studies demonstrate robust recovery under noise, diffusivity mismatch, and finite acquisition, and reveal a consistency--retention tradeoff governed by the manifold width. Photothermal experiments further confirm that the projection suppresses off-manifold spectral content while retaining the dominant thermal response. These results establish operator consistency as a spectral-selection principle for diffusive fields and provide a geometric framework for physics-informed processing of dissipative systems.
\end{abstract}

\maketitle

$\textit{Introduction}$---Diffusion occupies an unusual position among continuum transport processes~\cite{1,2,3}. Spatial disturbances spread and attenuate in a manner often described in terms of "diffusion waves," a framework developed to treat periodically driven thermal~\cite{4,5}, carrier-density~\cite{6,7}, photon-density~\cite{8,9}, mass-transport~\cite{10,11}, electromagnetic~\cite{12,13}, and related diffusive fields within a common mathematical description~\cite{14,15,16}. Yet the underlying dynamics are fundamentally different from those of a propagating hyperbolic wave: the diffusion equation is first order in time, its free spatial modes relax exponentially, and the complex wave numbers that arise under harmonic excitation describe simultaneous phase lag and attenuation rather than lossless propagation~\cite{17,18}. The language of waves is therefore useful but potentially deceptive. Diffusion lacks a conventional real-frequency dispersion relation of the type associated with conservative propagating media, while still possessing a highly structured spectrum imposed by its evolution operator~\cite{19}.
This distinction is not merely formal. Diffusive transport underlies a broad range of sensing and imaging modalities~\cite{20,21,22}. Thermal diffusion is exploited in photothermal radiometry~\cite{23}, modulated thermography~\cite{24}, thermal-wave microscopy~\cite{25}, and nondestructive testing~\cite{26,27,28}. Carrier-density diffusion waves provide access to electronic transport and recombination properties in semiconductors. Diffuse photon-density waves form the basis of frequency-domain optical spectroscopy and tomography in strongly scattering tissue. Electromagnetic diffusion governs eddy-current penetration and transient inspection of conducting structures~\cite{29,30}. Across these otherwise different applications, the same physical limitation repeatedly appears: diffusion progressively suppresses fine spatial information as it is transported from an excitation or buried inhomogeneity to the measurement surface. In thermography, this loss has been connected explicitly to entropy production and to depth-dependent limits on spatial resolution~\cite{31,32,33}. In diffuse optical imaging, increasing modulation frequency changes phase sensitivity but also increases attenuation of the measurable photon-density response~\cite{34}. Similar depth--resolution and bandwidth tradeoffs occur in electromagnetic and carrier-diffusion measurements. Thus the practical inverse problem is governed not only by measurement noise, but by the spectral selectivity inherent to diffusion itself.
Considerable effort has therefore been devoted to recovering information lost through diffusion. Photothermal and diffusion-wave methods have used phase-sensitive detection~\cite{35}, matched filtering~\cite{36}, and pulse compression~\cite{37} to improve depth localization and reject uncorrelated backgrounds. Inverse heat-conduction and thermographic reconstruction methods introduce regularization or prior information to stabilize the exponentially ill-conditioned recovery of subsurface structure. The virtual-wave concept takes a complementary route by transforming measured diffusive data into an auxiliary wave field, allowing reconstruction tools developed for reversible wave propagation to be applied after a regularized inversion of the diffusion process~\cite{38,39}. More recently, physics-informed optimization and learning approaches have incorporated PDE residuals, boundary conditions, or learned priors directly into inverse problems governed by parabolic equations~\cite{40,41}. These methods demonstrate the value of embedding physical knowledge in reconstruction, but their physical constraints are generally imposed in the image, parameter, or solution space rather than used to define a selection geometry for the measured spectrum itself.
A second line of work has emphasized the spectral and geometric structure generated by diffusion operators. Spatial Fourier decomposition diagonalizes the homogeneous diffusion generator, associating each spatial scale with a definite exponential decay rate. Operator-theoretic descriptions of relaxation dynamics similarly characterize dissipative systems through decaying eigenmodes and eigenfunctionals rather than propagating oscillatory modes. In data analysis, diffusion maps and diffusion wavelets exploit spectra of Markov~\cite{42} or diffusion operators to construct intrinsic coordinates, multiscale geometries, and compressed representations~\cite{43}. These developments connect diffusion, operator spectra, and geometry, but they do not use the governing spectral relation to assess individual components of a measured physical field. This distinction is important because conventional spectral processing typically selects components by frequency, wavenumber, or magnitude. Although diffusion strongly attenuates high spatial frequencies, attenuation does not imply physical inadmissibility: a weak high-wavenumber component may satisfy the diffusion dynamics exactly, whereas a stronger low-wavenumber component may violate them. Magnitude-based filtering therefore conflates spectral scale with operator consistency and may suppress physically valid fine-scale information.
This motivates a different criterion: rather than selecting components by spectral magnitude, we ask whether they are dynamically admissible. For diffusion, this means identifying the joint relation between spatial structure and temporal relaxation and using departure from that relation as a spectral coordinate. The central question is therefore whether the dissipative spectrum defines an operator-imposed admissible set that can serve directly as a physics-based spectral-selection principle.

For a homogeneous isotropic medium,
\begin{equation}
	\frac{\partial T}{\partial t}
	=
	\alpha\nabla^2T,
	\label{eq:diffusion}
\end{equation}
where $T(\mathbf{x},t)$ is the temperature field and $\alpha$ is the thermal diffusivity. Spatial Fourier decomposition diagonalizes the diffusion generator. Writing
\begin{equation}
	T(\mathbf{x},t)
	=
	\int
	A(\mathbf{k})
	\exp(i\mathbf{k}\cdot\mathbf{x})
	\exp[\lambda(\mathbf{k})t]
	\,d\mathbf{k},
	\label{eq:modal_decomposition}
\end{equation}
substitution into Eq.~\eqref{eq:diffusion} gives the generator eigenvalue
$
	\lambda(\mathbf{k})
	=
	-\alpha |\mathbf{k}|^2.
$
It is convenient to introduce the positive modal decay rate
$\eta\equiv-\lambda$, so that
$
	\eta
	=
	\alpha |\mathbf{k}|^2.
$
Here $\eta$ is a decay-rate coordinate associated with the spectrum of the diffusion generator, rather than a temporal Fourier frequency. Equivalently, it corresponds to the positive decay coordinate obtained from the Laplace-domain pole condition of the diffusion operator.
The admissible diffusion modes therefore occupy the operator-defined set
\begin{equation}
	\mathcal{M}_d
	=
	\left\{
	(\mathbf{k},\eta):
	\eta=\alpha|\mathbf{k}|^2
	\right\},
	\label{eq:manifold}
\end{equation}
which forms a parabola in $(k,\eta)$ and a paraboloid in $(k_x,k_y,\eta)$. We refer to $\mathcal{M}_d$ as the diffusion spectral manifold.
For experimentally sampled fields, the decay-rate spectrum is obtained from the measured temporal evolution rather than from a temporal Fourier transform. We first perform a spatial Fourier transform,
$
T(\mathbf{x},t)\rightarrow \widetilde{T}(\mathbf{k},t),
$
and represent each spatial mode as a superposition of exponential relaxation modes,
$
\widetilde{T}(\mathbf{k},t)
=
\int_{0}^{\infty}
S(\mathbf{k},\eta)e^{-\eta t}\,d\eta.
$
For discrete measurements $t_n$ and a sampled decay-rate grid $\eta_j$, this relation becomes
$
\widetilde{\mathbf{T}}_{\mathbf{k}}
=
\mathbf{K}\mathbf{s}_{\mathbf{k}},
K_{nj}
=
\exp(-\eta_j t_n).
$
Because recovery of the decay-rate distribution constitutes an inverse-Laplace-type problem, $S(\mathbf{k},\eta)$ is estimated using regularized nonnegative least squares,
\begin{equation}
\mathbf{s}_{\mathbf{k}}
=
\underset{\mathbf{s}\geq 0}{\operatorname{arg\,min}}
\left(
\left\|
\mathbf{K}\mathbf{s}
-
\widetilde{\mathbf{T}}_{\mathbf{k}}
\right\|_2^2
+
\lambda_{\eta}
\left\|
\mathbf{L}\mathbf{s}
\right\|_2^2
\right),
\end{equation}
where $\mathbf{L}$ is a discrete smoothness operator and $\lambda_{\eta}$ controls regularization. The resulting $S(\mathbf{k},\eta)$ is therefore an estimated modal-decay representation of the measured field, rather than a conventional temporal Fourier spectrum.
Real measurements and finite-resolution modal representations need not remain concentrated on an infinitesimally thin manifold. Noise, finite acquisition, parameter uncertainty, boundary effects, and source superposition can distribute spectral energy away from the ideal operator relation. We therefore define the diffusion-operator residual
$
	r_d(\mathbf{k},\eta)
	=
	\eta-\alpha|\mathbf{k}|^2,
$
and its magnitude
$
	d_d(\mathbf{k},\eta)
	=
	\left|r_d(\mathbf{k},\eta)\right|.
$
The quantity $d_d$ measures violation of the diffusion dispersion relation along the decay-rate coordinate; it should not be interpreted as the Euclidean shortest distance to the paraboloidal surface.
A finite-width soft projection is then defined by
\begin{equation}
	W_{\sigma}(\mathbf{k},\eta)
	=
	\exp\!\left[
	-\frac{r_d^2(\mathbf{k},\eta)}
	{2\sigma^2}
	\right],
	\label{eq:weight}
\end{equation}
with
$
	S_{\mathrm{proj}}
	=
	W_{\sigma}S_{\mathrm{meas}}.
$
Here $\sigma$ has the same units as the decay rate and determines the tolerance to operator inconsistency. We use the term ``soft projection'' because Eq.~\eqref{eq:weight} is a smooth finite-width relaxation of an ideal manifold selection rather than an idempotent orthogonal projection. The essential distinction is that spectral transmission is controlled by compatibility with the governing operator, rather than by $|\mathbf{k}|$ or $\eta$ separately.

\begin{figure*}[t]
  \centering
  \includegraphics[width=\textwidth]{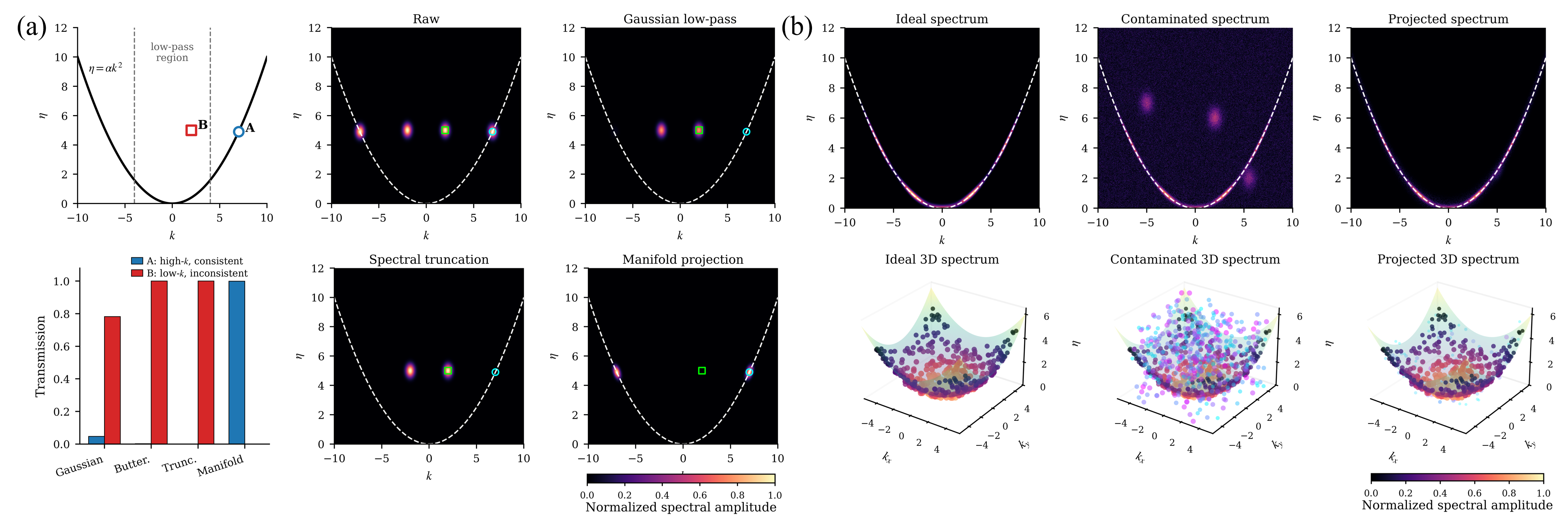}
  \caption{Operator-consistency selection and dispersion-constrained spectral recovery. (a) Comparison between conventional spectral-magnitude filtering and operator-consistency selection. A high-wavenumber mode A satisfies the diffusion relation $\eta=\alpha k^2$, whereas the lower-wavenumber mode B has a large diffusion-operator residual. Gaussian low-pass filtering and spectral truncation preferentially retain B, while the dispersion-constrained projection preserves A and suppresses B; the corresponding transmission coefficients summarize the inequivalent spectral ordering. (b) Direct visualization of the projection mechanism. An ideal spectrum concentrated near the diffusion manifold is contaminated by broadband and localized off-manifold components and subsequently recovered by the finite-width soft projection. The lower row shows the corresponding extension to two spatial dimensions, where the admissible modes lie near the paraboloidal manifold $\eta=\alpha(k_x^2+k_y^2)$.}
  \label{fig1}
\end{figure*}
\begin{table}[t]
	\caption{\label{tab:sample_exp}
		Physical properties of the stainless-steel specimen, defect geometry, and experimental parameters used for photothermal measurements.}
	\begin{ruledtabular}
		\begin{tabular}{lll}
			Parameter & Symbol & Value \\
			\hline
			\multicolumn{3}{c}{\textit{Specimen}} \\
			Material & -- & 316L stainless steel \\
			Lateral size & $d_{\mathrm{s}}$ & $58.5\times58.5~\mathrm{mm}^2$ \\
			Thickness & $L$ & $4.5~\mathrm{mm}$ \\
			Thermal diffusivity & $\alpha$ & $3.76~\mathrm{mm^2\,s^{-1}}$ \\
			Thermal conductivity & $k$ & $15~\mathrm{W\,m^{-1}K^{-1}}$ \\
			Density & $\rho$ & $7950~\mathrm{kg\,m^{-3}}$ \\
			Specific heat capacity & $c_p$ & $502~\mathrm{J\,kg^{-1}K^{-1}}$ \\[2pt]
			
			\multicolumn{3}{c}{\textit{Defects}} \\
			Defect size & $d_{\mathrm{def}}$ & $2\times2~\mathrm{mm}^2$ \\
			Defect depth & $L_{\mathrm{def}}$ & $0.5~\mathrm{mm}$ \\
			Separation & $d_{\mathrm{sep}}$ & $0.5,\ 1,\ 2,\ 4~\mathrm{mm}$ \\
			Contrast & $\zeta$ & $\approx0.494$ \\[2pt]
			
			\multicolumn{3}{c}{\textit{Scanning and illumination}} \\
			Region of interest & -- & $43\times4.3~\mathrm{mm}^2$ \\
			No. of measurements & $n_m$ & $403$ \\
			Grid spacing & $r_d$ & $0.743~\mathrm{mm}$ \\
			Laser power & $\hat{Q}$ & $15~\mathrm{W}$ \\
			Laser spot size & $d_{\mathrm{spot}}$ & $0.6~\mathrm{mm}$ \\
			Pulse duration & $t_{\mathrm{pulse}}$ & $500~\mathrm{ms}$ \\[2pt]
			
			\multicolumn{3}{c}{\textit{Infrared acquisition}} \\
			Spatial resolution & $\Delta x,\Delta y$ & $52~\mu\mathrm{m}$ \\
			Sampling frequency & $f_{\mathrm{cam}}$ & $100~\mathrm{Hz}$ \\
			Spectral band & -- & MWIR \\
		\end{tabular}
	\end{ruledtabular}
\end{table}

Figure~\ref{fig1}(a) provides a direct counterexample to any identification of the proposed projection with conventional low-pass filtering. Consider two components satisfying
$
|k_A|>|k_B|,
\qquad
d_d(A)<d_d(B).
$
Any monotonic low-pass rule based only on $|k|$ preferentially transmits B over A, whereas an operator-consistency rule based on $d_d$ preferentially transmits A over B. The two criteria therefore induce fundamentally inequivalent spectral orderings. In the construction shown in Fig.~\ref{fig1}(a), mode A satisfies $\eta_A=\alpha k_A^2$ and is operator-consistent despite its larger wavenumber, whereas mode B has a smaller wavenumber but violates the diffusion relation. Gaussian, Butterworth, and hard-cutoff filtering consequently favor B, while the dispersion-constrained soft projection preserves A and suppresses B.
Figure~\ref{fig1}(b) exposes the corresponding projection mechanism directly. The ideal modal spectrum is concentrated near $r_d=0$, whereas broadband and localized operator-inconsistent contamination distributes spectral energy away from the diffusion manifold. Multiplication by Eq.~\eqref{eq:weight} suppresses components with large diffusion residual while preserving those compatible with the governing operator, without imposing an independent cutoff in $k$ or $\eta$. The same geometric construction extends naturally to two spatial dimensions, for which
$
\eta
=
\alpha
\left(
k_x^2+k_y^2
\right),
$
and
$
r_d
=
\eta-\alpha(k_x^2+k_y^2).
$
The admissible modes therefore occupy a paraboloidal manifold in $(k_x,k_y,\eta)$ space, demonstrating that the operator-consistency principle is independent of spatial dimension.
The comparison in Fig.~\ref{fig1} is intended to establish the inequivalence of magnitude- and operator-based spectral selection rather than to constitute a general performance benchmark. A quantitative comparison with operator-residual Tikhonov regularization based on the same diffusion physics is provided in the Supplemental Material.

We next test whether this geometric criterion remains useful when the ideal manifold is obscured by measurement imperfections. The dispersion residual
\begin{equation}
	E_{\mathrm{disp}}(S)
	=
	\frac{
		\left\|
		r_d(k,\eta)S(k,\eta)
		\right\|_2
	}{
		\left\|
		S(k,\eta)
		\right\|_2
	},
	\label{eq:edisp}
\end{equation}
quantifies the energy-weighted violation of the diffusion-operator relation. Under additive noise, off-manifold contamination increases $E_{\mathrm{disp}}$, whereas 
$
S_{\mathrm{proj}}
=
W_{\sigma}S_{\mathrm{meas}}.
$ concentrates spectral energy back toward $\mathcal{M}_d$. At an input SNR of $10\,\mathrm{dB}$, the PSNR increases from $29.46$ to $38.97\,\mathrm{dB}$ and $E_{\mathrm{disp}}$ decreases from $1.601$ to $0.116$. The projected spectrum remains close to the diffusion geometry over the investigated noise range, indicating recovery of operator consistency rather than spectral smoothing alone.
\begin{figure*}[t]
  \centering
  \includegraphics[width=\textwidth]{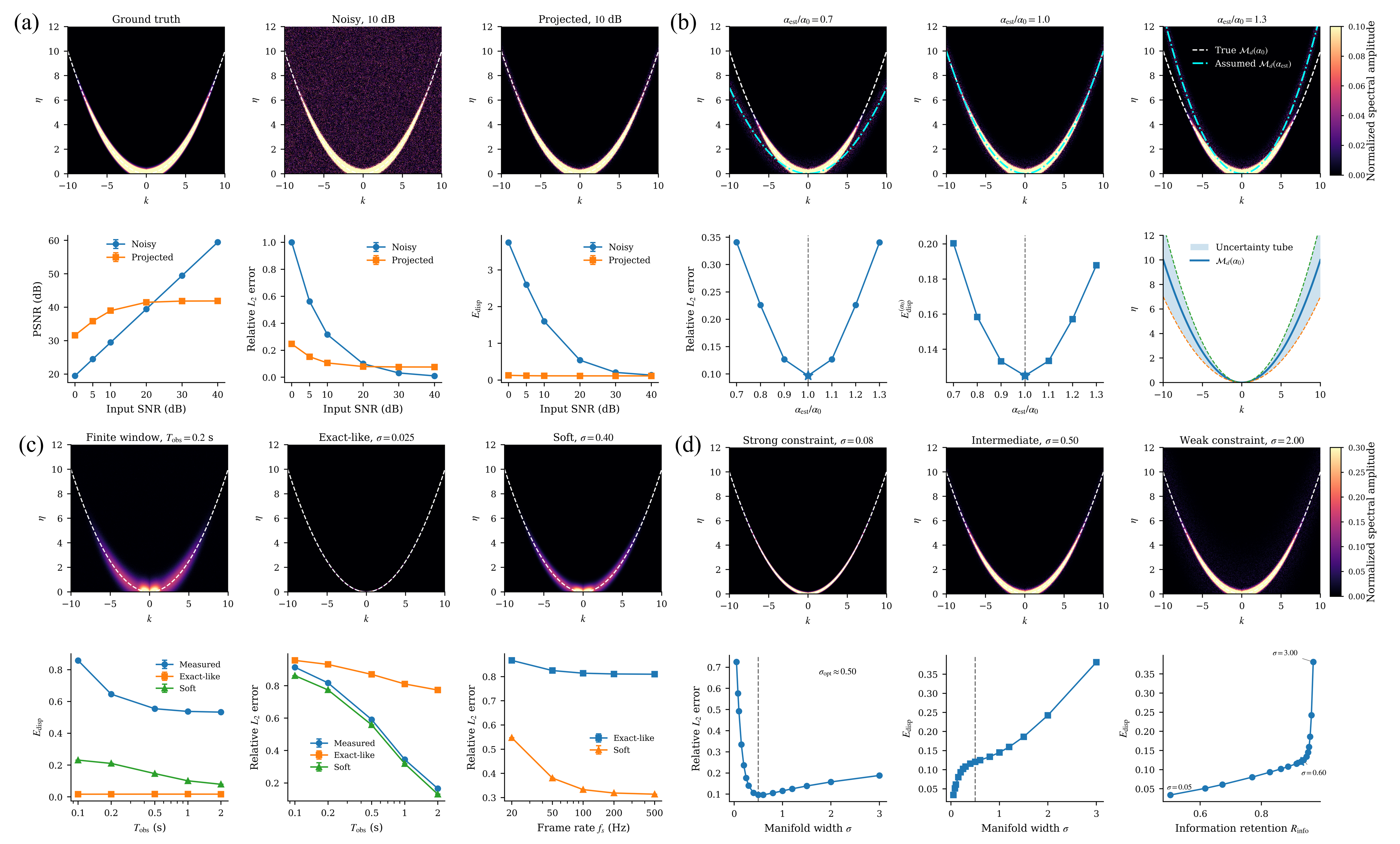}
  \caption{Robustness and finite-width character of dispersion-constrained soft projection. (a) Noise robustness. (b) Sensitivity to diffusivity mismatch. (c) Finite-acquisition effects. (d) Consistency--retention tradeoff controlled by the width $\sigma$. A nonzero width accommodates finite-resolution and parameter-induced deviations from the ideal operator relation while rejecting strongly inconsistent components.}
  \label{fig2}
\end{figure*}
Figure~\ref{fig2} further examines the robustness and practical limits of the dispersion-constrained projection. As shown in Fig.~\ref{fig2}(a), the projection consistently improves spectral recovery over the investigated noise range, demonstrating that the operator constraint remains effective even when broadband contamination becomes substantial. Figure~\ref{fig2}(b) shows that the reconstruction is also tolerant to moderate errors in the assumed diffusivity: although parameter mismatch shifts the nominal diffusion manifold, the physically relevant spectral components remain recoverable within a finite neighborhood of the ideal operator relation. The finite-acquisition study in Fig.~\ref{fig2}(c) further indicates that the method remains effective under limited observation windows, where spectral broadening prevents the measured field from lying exactly on the theoretical manifold. Finally, Fig.~\ref{fig2}(d) reveals a clear consistency--retention tradeoff with respect to the manifold width $\sigma$, with the best reconstruction occurring at a finite rather than vanishing width.
The finite width $\sigma$ is not merely a numerical tuning parameter, but sets the tolerated residual from the ideal diffusion-operator relation. In the limit $\sigma\rightarrow0$, the weighting approaches an increasingly restrictive selection of modes satisfying $r_d=0$. A finite value of $\sigma$ instead accommodates deviations arising from finite spectral resolution, noise, and uncertainty in the assumed material parameters. Increasing $\sigma$ therefore produces a consistency--retention tradeoff: excessively small values reject recoverable information, whereas excessively large values admit progressively more operator-inconsistent content. The resulting optimum occurs at a finite width. Moderate diffusivity errors likewise shift the assumed manifold without immediately destroying recoverability because physically relevant components can remain inside the finite admissible neighborhood. Detailed parameter sweeps and information-retention analyses are provided in the Supplemental Material.

The central consequence is that a dissipative field can possess a structured operator spectrum even in the absence of a propagating-wave dispersion relation. For diffusion, the generator spectrum separates two notions that conventional filtering conflates: spectral scale and operator consistency. A high-wavenumber component need not be unphysical, and a low-wavenumber component need not satisfy the governing dynamics. The appropriate selection coordinate is therefore not necessarily frequency magnitude, but the residual from the operator-defined admissible set. In this sense, the transformation
$
	(k,\eta)
	\longrightarrow
	\left(k,r_d\right)
$
introduces an operator-consistency coordinate for spectral analysis. The finite-width soft projection considered here provides one realization of this principle. More broadly, whenever the spectrum of a dissipative or non-Hermitian evolution operator restricts admissible modes to a structured subset of spectral space, residuals from that operator relation can provide physics-informed coordinates for reconstruction, denoising, and inverse analysis.
\begin{figure*}[t]
	\centering
	\includegraphics[width=0.8\textwidth]{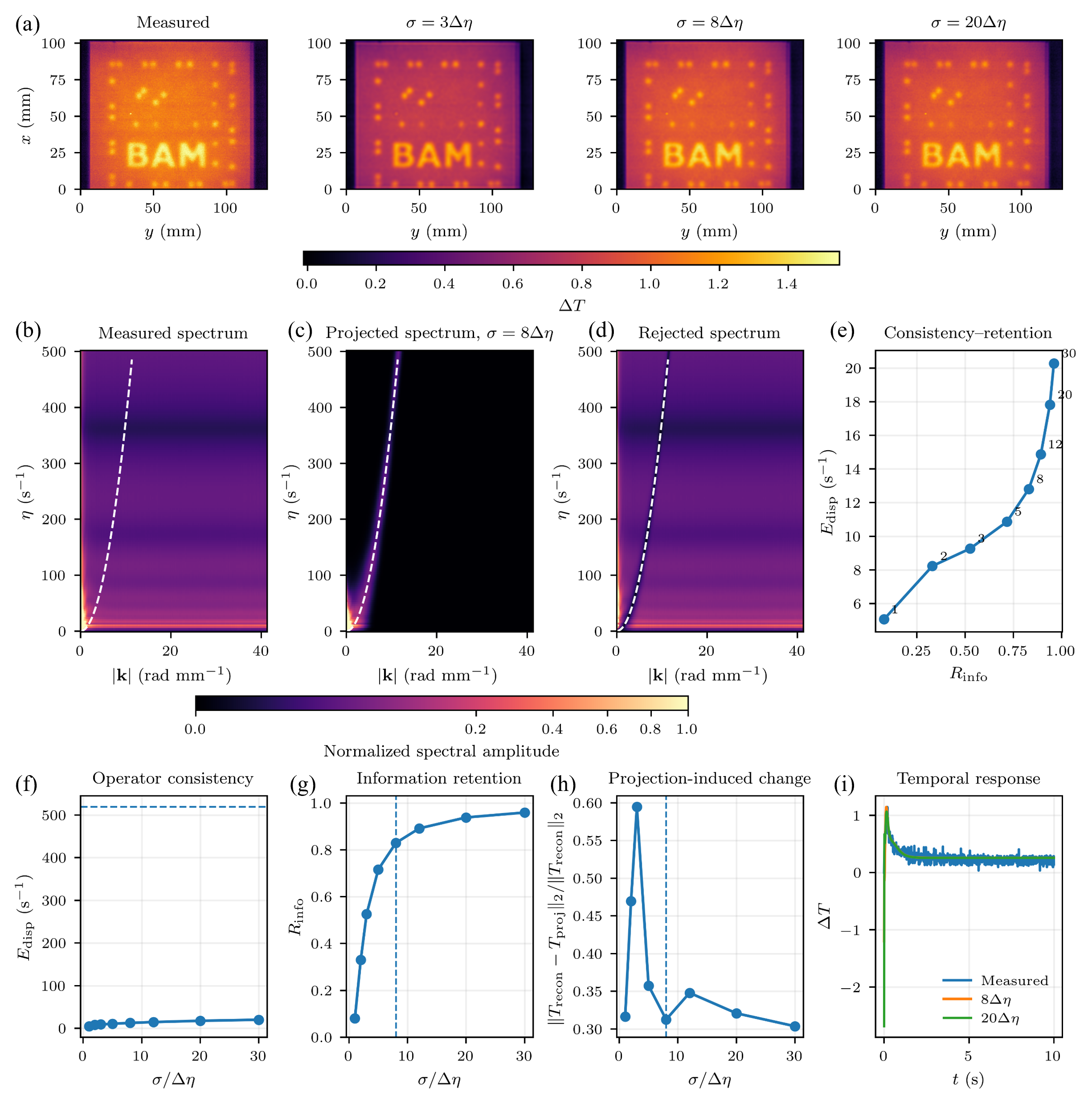}
	\caption{Experimental validation of dispersion-constrained manifold projection.
		(a) Measured temperature field and projected fields for $\sigma=3\Delta\eta$, $8\Delta\eta$, and $20\Delta\eta$. (b–d) Measured, projected, and rejected spectra; the white dashed curve denotes the theoretical diffusion manifold $\eta=\alpha |\mathbf{k}|^2$. (e) Tradeoff between dispersion consistency and information retention, with numbers indicating $\sigma/\Delta\eta$. (f) Dispersion residual $E_{\mathrm{disp}}$, (g) information-retention ratio $R_{\mathrm{info}}$, and (h) projection-induced relative change as functions of $\sigma/\Delta\eta$. The dashed lines indicate the unprojected reference in (f) and the representative choice $\sigma=8\Delta\eta$ in (g,h). (i) Representative temporal responses before and after projection. The results demonstrate that finite-width manifold projection suppresses off-manifold spectral content while retaining the dominant physically consistent thermal response.}
	\label{fig3}
\end{figure*}
\begin{table}[t]
	\caption{SNR performance of the dispersion-constrained projection for different normalized manifold widths $\sigma/\Delta\eta$. The SNR gain is calculated relative to the original field.}
	\label{tab:snr_comparison}
	\centering
	\begin{ruledtabular}
		\begin{tabular}{lccc}
			Method & $\sigma/\Delta\eta$ & Peak SNR (dB) & SNR gain (dB) \\
			\hline
			Original  & -- & 9.65  & 0.00  \\
			Projected & 1  & 24.19 & 14.55 \\
			Projected & 2  & \textbf{24.49} & \textbf{14.85} \\
			Projected & 3  & 23.79 & 14.14 \\
			Projected & 5  & 23.38 & 13.73 \\
			Projected & 8  & 23.13 & 13.48 \\
			Projected & 12 & 22.94 & 13.29 \\
			Projected & 20 & 22.65 & 13.00 \\
			Projected & 30 & 22.35 & 12.70 \\
		\end{tabular}
	\end{ruledtabular}
\end{table}
We finally validate the dispersion-constrained projection experimentally using photothermal measurements on a 316L stainless-steel specimen containing subsurface defects~\cite{44}; the specimen properties and acquisition parameters are summarized in Table~\ref{tab:sample_exp}. Figure~\ref{fig3}(a) shows that the finite-width projection preserves the dominant defect contrast while progressively approaching the measured field as $\sigma$ is increased. More importantly, the spectral representation in Figs.~\ref{fig3}(b)--(d) reveals that the retained energy is concentrated around the theoretical diffusion manifold $\eta=\alpha|\mathbf{k}|^2$, whereas the rejected spectrum is predominantly distributed away from it. This directly demonstrates that the projection acts according to operator consistency rather than spectral magnitude alone.
The experimental data further exhibit the same consistency--retention tradeoff predicted numerically. As shown in Figs.~\ref{fig3}(e)--(h), increasing $\sigma$ raises the information-retention ratio $R_{\mathrm{info}}$ but simultaneously admits progressively more dispersion-inconsistent content, leading to an increase in $E_{\mathrm{disp}}$. A representative intermediate width, $\sigma=8\Delta\eta$, retains more than $80\%$ of the measured spectral energy while maintaining a substantially smaller dispersion residual than the unprojected measurement. The corresponding temporal response in Fig.~\ref{fig3}(i) remains close to the measured signal, confirming that the projection primarily removes operator-inconsistent contributions rather than altering the underlying thermal dynamics. These observations experimentally support the interpretation of $\sigma$ as a finite geometric tolerance to the diffusion manifold and reinforce the need for a soft, rather than zero-width, projection.
\begin{figure*}[t]
	\centering
	\includegraphics[width=\textwidth]{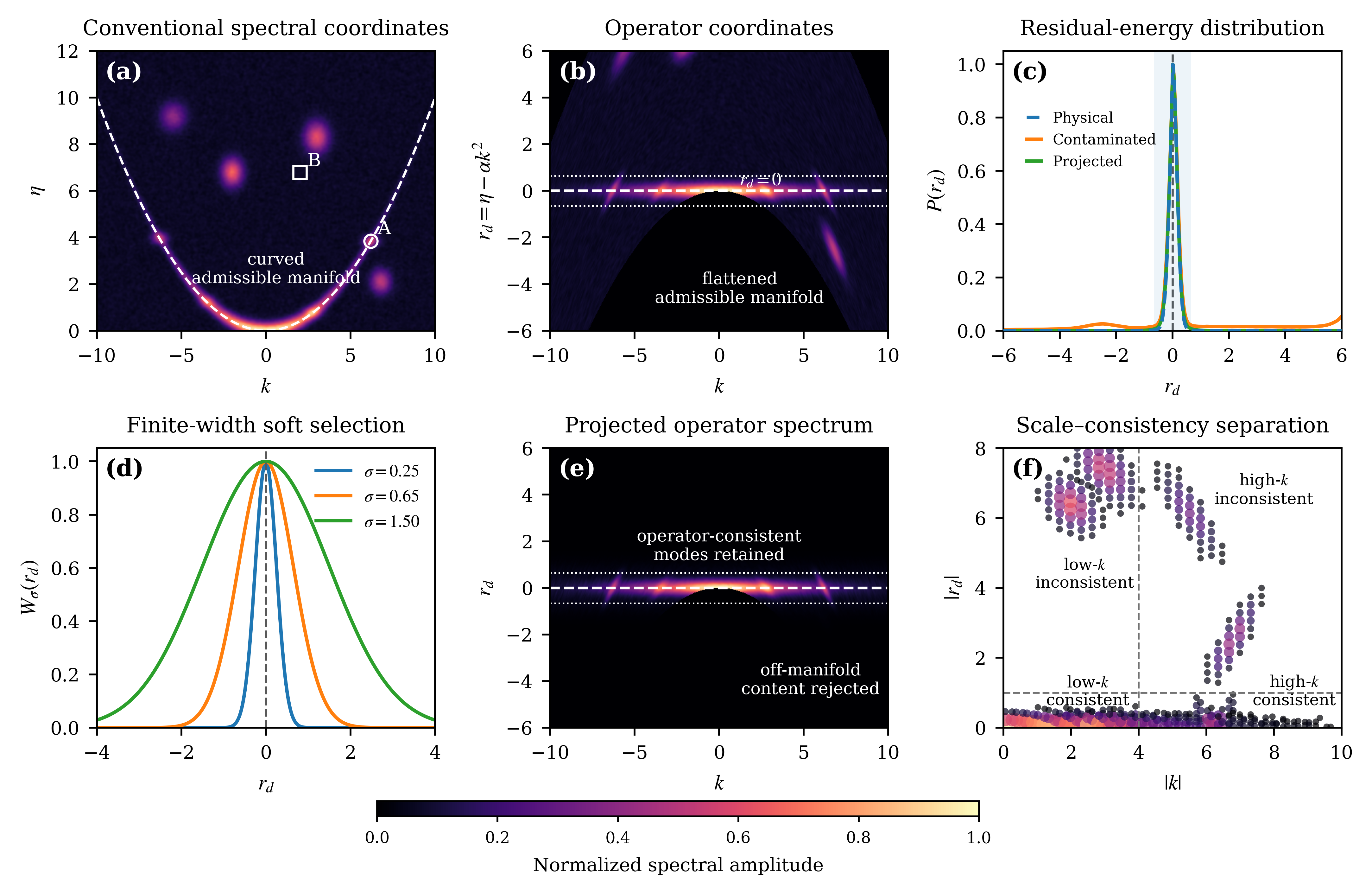}
	\caption{Operator-coordinate representation and dispersion-constrained spectral projection.
		(a) Measured spectrum in the conventional $(k,\eta)$ coordinates, where the diffusion-consistent states lie near the curved manifold $\eta=\alpha k^2$ (white dashed curve). Points A and B illustrate, respectively, a high-$k$ manifold-consistent component and a lower-$k$ off-manifold component.
		(b) Transformation to the operator residual $r_d=\eta-\alpha k^2$ flattens the diffusion manifold onto $r_d=0$, converting dispersion consistency into distance from a horizontal axis. The dotted lines indicate the representative finite-width neighborhood $|r_d|=\sigma$.
		(c) Normalized residual-energy distributions $P(r_d)=\int |S(k,r_d)|^2,dk$ for the physical, contaminated, and projected spectra. Off-manifold contamination broadens the residual distribution, whereas projection concentrates the spectral energy around $r_d=0$.
		(d) Gaussian soft-selection functions $W_\sigma(r_d)=\exp[-r_d^2/(2\sigma^2)]$ for representative manifold widths, illustrating the continuous tradeoff between dispersion selectivity and spectral retention.
		(e) Projected spectrum for $\sigma=0.65$, showing preferential retention of operator-consistent modes and suppression of off-manifold content.
		(f) Spectral amplitude represented in the $(|k|,|r_d|)$ plane, separating spectral scale from dispersion consistency. The dashed partitions are illustrative and are not used by the projection. Both low- and high-$k$ components can therefore be retained when dispersion-consistent, whereas inconsistent components can be rejected independently of their spectral scale.}
	\label{fig4}
\end{figure*}
As an independent measure of experimental signal quality, we further evaluate the signal-to-noise ratio (SNR) directly from the reconstructed temperature fields. At each time instant, the thermal SNR is defined as
\begin{equation}
	\mathrm{SNR}(t)
	=
	20\log_{10}
	\left[
	\frac{
		\left|
		\mu_{\mathrm{sig}}(t)-\mu_{\mathrm{bg}}(t)
		\right|
	}{
		\sigma_{\mathrm{bg}}(t)
	}
	\right],
	\label{eq:thermal_snr}
\end{equation}
where $\mu_{\mathrm{sig}}(t)$ and $\mu_{\mathrm{bg}}(t)$ denote the mean temperatures within fixed defect and defect-free background regions, respectively, and $\sigma_{\mathrm{bg}}(t)$ is the spatial standard deviation of the background region. The peak SNR is then defined as
$
	\mathrm{SNR}_{\mathrm{peak}}
	=
	\max_{t}\mathrm{SNR}(t).
$
The unprojected field exhibits a peak SNR of $9.65~\mathrm{dB}$ at $t=0.30~\mathrm{s}$, as shown in Table~\ref{tab:snr_comparison}. Dispersion-constrained projection substantially increases the peak SNR over the entire investigated range of manifold widths, yielding values between $22.35$ and $24.49~\mathrm{dB}$. The maximum SNR of $24.49~\mathrm{dB}$ is obtained at $\sigma=2\Delta\eta$, corresponding to an improvement of $14.85~\mathrm{dB}$ relative to the unprojected measurement. As the manifold width is increased beyond this value, the SNR decreases gradually, consistent with the progressive admission of off-manifold spectral content.
Importantly, the SNR is evaluated directly from the reconstructed temperature field and does not involve the dispersion residual $r_d$ or $E_{\mathrm{disp}}$. It therefore provides an independent experimental measure showing that enforcement of operator consistency is accompanied by a substantial improvement in thermal defect detectability rather than merely a reduction of the operator-residual metric.

The geometric interpretation of the projection is illustrated in Fig.~\ref{fig4}. In the conventional spectral coordinates $(k,\eta)$, diffusion-consistent components occupy the curved dispersion manifold $\eta=\alpha k^2$ (Fig.~\ref{fig4}(a)). This representation entangles spectral scale with physical admissibility: a component may occur at large $|k|$ while remaining fully consistent with diffusion dynamics, whereas a low-$|k|$ component may lie far from the admissible manifold. Consequently, spectral magnitude alone does not provide a physically meaningful criterion for separating admissible and nonadmissible content.
We therefore introduce the operator residual
$
	r_d=\eta-\alpha k^2,
$
which measures departure from the diffusion dispersion relation. Under the coordinate transformation $(k,\eta)\rightarrow(k,r_d)$, the curved physical manifold is flattened onto the line $r_d=0$ (Fig.~\ref{fig4}(b)). Dispersion consistency is thus converted into a transverse distance in operator space, while the coordinate $k$ remains available to describe spectral scale. This separation is directly visible in the residual-energy distribution
$
	P(r_d)=\int |S(k,r_d)|^2,dk ,
$
for which physically consistent energy is concentrated near $r_d=0$, whereas off-manifold contamination produces finite-residual contributions (Fig.~\ref{fig4}(c)).
Rather than imposing an exact manifold constraint, we employ the finite-width projection
$
	W_{\sigma}(r_d)
	=
	\exp \left(-r_d^2/(2\sigma^2)\right),
	S_{\mathrm{proj}}(k,r_d)
	=
	W_{\sigma}(r_d)S_{\mathrm{meas}}(k,r_d),
$
where $\sigma$ controls the admissible neighborhood around the dispersion manifold (Fig.~\ref{fig4}(d)). The resulting spectrum remains concentrated around $r_d=0$ while off-manifold contributions are continuously attenuated rather than abruptly truncated (Fig.~\ref{fig4}(e)).
Importantly, the projection criterion depends on operator consistency rather than spectral magnitude. Figure~\ref{fig4}(f) makes this distinction explicit by separating the spectral scale $|k|$ from the residual $|r_d|$. Both low- and high-$k$ modes can be physically admissible when $|r_d|$ is small, while components at either scale can be inconsistent when $|r_d|$ is large. The operator-coordinate projection therefore acts as a dispersion-selective filter rather than a conventional low-pass filter, preserving high-spatial-frequency information whenever it remains compatible with the governing diffusion operator.

\begin{figure*}[t]
	\centering
	\includegraphics[width=0.8\textwidth]{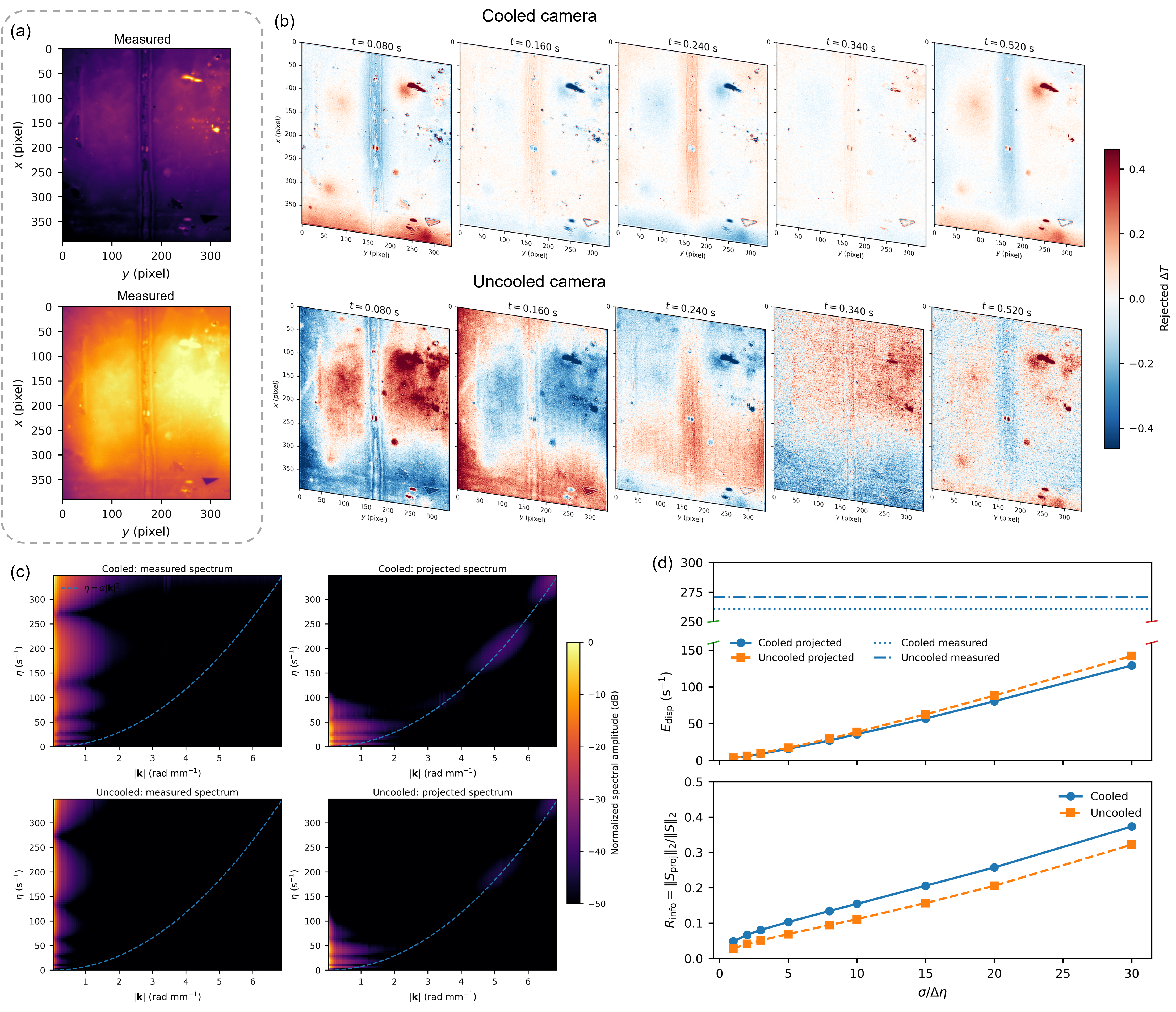}
	\caption{
		Effect of detector cooling on dispersion-constrained manifold projection.
		(a) Measured fields obtained with cooled (top) and uncooled (bottom) infrared cameras.
		(b) Rejected components, $\Delta T_{\mathrm{rej}}=T_{\mathrm{meas}}-T_{\mathrm{proj}}$, at representative times.
		(c) Corresponding measured and projected spectra in $(|\mathbf{k}|,\eta)$ space; the cyan dashed curve denotes the diffusion manifold $\eta=\alpha|\mathbf{k}|^{2}$.
		(d) Dispersion residual $E_{\mathrm{disp}}$ (top) and information-retention ratio $R_{\mathrm{info}}$ (bottom) versus $\sigma/\Delta\eta$. The uncooled measurement contains substantially stronger off-manifold contributions, while projection drives both data sets toward operator consistency with increasing spectral retention as the manifold width is relaxed.
	}
	\label{fig5}
\end{figure*}

We further tested whether the operator-consistency criterion remains
effective when the measurement quality is substantially degraded.
Figure~\ref{fig5} compares data acquired using cooled and uncooled infrared
cameras. This data is available as open source at \cite{45}. Although the dominant thermal structures are recovered in both
measurements (Fig.~\ref{fig5}(a)), the rejected fields in Fig.~\ref{fig5}(b) reveal a
markedly stronger and more spatially distributed contribution for the
uncooled detector. The spectral representations in Fig.~\ref{fig5}(c) show that
this additional content is predominantly distributed away from the
reference diffusion manifold $\eta=\alpha|\mathbf{k}|^{2}$, whereas the
cooled measurement is more strongly concentrated within its finite
neighborhood. Dispersion-constrained projection suppresses these
off-manifold contributions in both cases without imposing an independent
cutoff in $|\mathbf{k}|$.
Figure~\ref{fig5}(d) quantifies this behavior. The projected
$E_{\mathrm{disp}}$ remains substantially below the corresponding
unprojected value for both cameras over the investigated
$\sigma/\Delta\eta$ range. Increasing the manifold width simultaneously
raises $R_{\mathrm{info}}$ and $E_{\mathrm{disp}}$, reproducing the
consistency--retention tradeoff associated with finite-width projection.
At the same $\sigma/\Delta\eta$, the cooled data retain a larger fraction
of the measured spectrum, consistent with their lower off-manifold
contamination. The comparison therefore shows that increased detector
noise manifests primarily as operator-inconsistent spectral content and
can be selectively suppressed by the dispersion-constrained projection.

$\textit{Conclusion}$---We have shown that diffusion defines not only a decay process but also an operator-imposed spectral geometry. By introducing the residual $r_d=\eta-\alpha|\mathbf{k}|^2$, the curved diffusion manifold is transformed into an operator-consistency coordinate that separates physical admissibility from spectral scale. This distinction leads naturally to a finite-width soft projection that suppresses dispersion-inconsistent components without imposing an independent cutoff in wavenumber or decay rate. Numerical results establish robustness against noise, parameter mismatch, and finite acquisition, while revealing the manifold width $\sigma$ as a geometric regularization parameter controlling the tradeoff between operator consistency and information retention. Photothermal experiments further confirm that off-manifold spectral content can be selectively suppressed while preserving the dominant thermal response. More broadly, the results suggest a shift from frequency-based to operator-based spectral selection: whenever the governing evolution operator confines physically admissible states to a structured subset of spectral space, distance from that set can provide a physics-informed coordinate for filtering, reconstruction, and inverse analysis. This perspective may extend dispersion-constrained spectral processing beyond thermal diffusion to a broader class of dissipative and non-Hermitian systems.

\textit{Acknowledgments}---This work was supported by the Adolf Martens Fellowship (Grant No.~BAM-AMF-2025-1).
\begin{figure}[tb]
	\centering
	\includegraphics[width=0.45\textwidth]{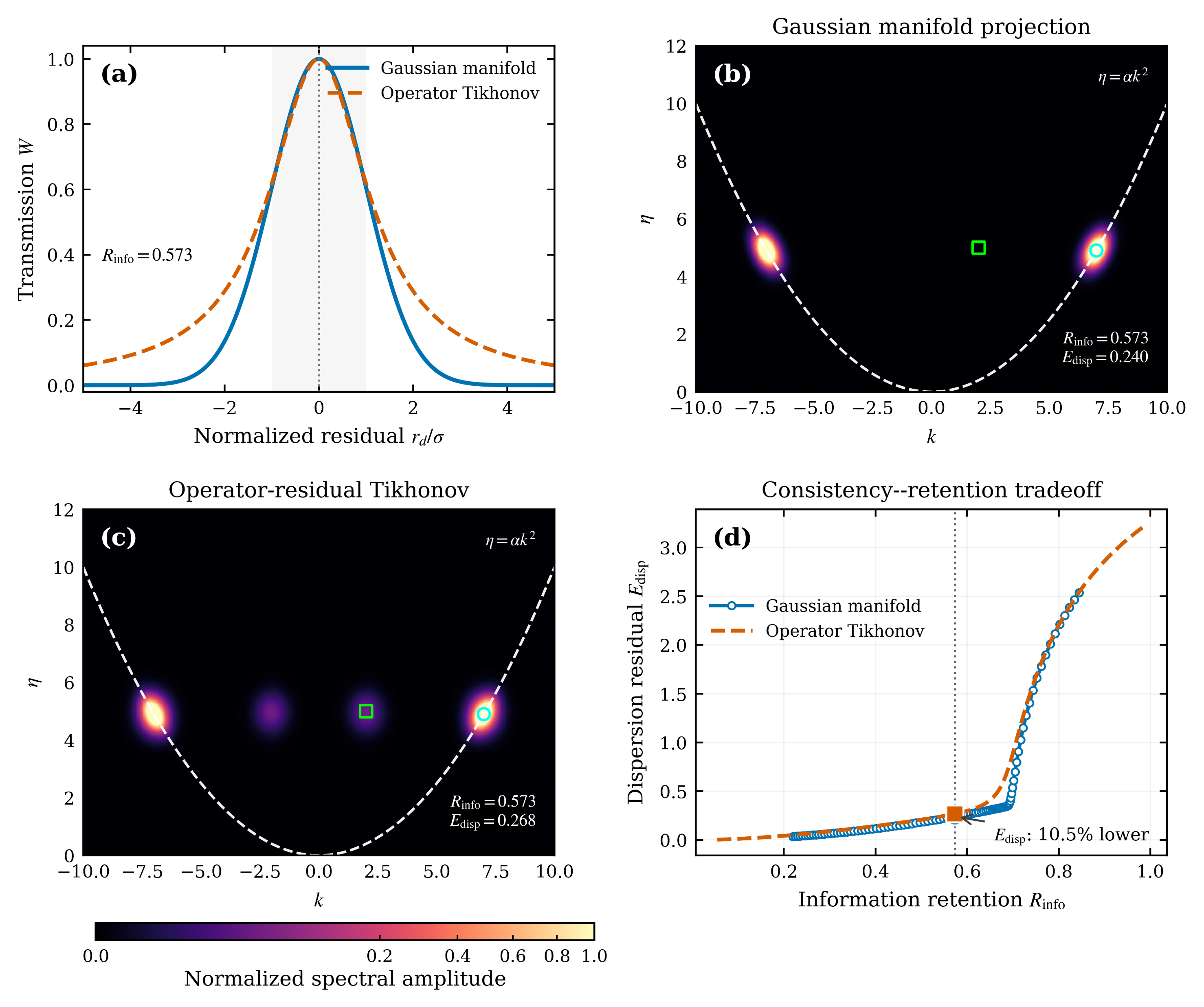}
	\caption{Comparison between the proposed finite-width Gaussian manifold projection
		and operator-residual Tikhonov regularization.
		(a) Transmission functions expressed in the normalized operator-residual
		coordinate $r_d/\sigma$. The Gaussian weighting decays exponentially away
		from the diffusion manifold, whereas the Tikhonov weighting exhibits
		broader algebraic tails. The two filters are compared at matched information
		retention, $R_{\mathrm{info}}=0.573$.
		(b) Spectrum obtained using the Gaussian manifold projection.
		The high-$k$ component lying on the reference diffusion manifold
		$\eta=\alpha k^2$ is retained, while the off-manifold component is strongly
		suppressed.
		(c) Spectrum obtained using operator-residual Tikhonov regularization at the
		same information-retention level. The physically consistent component is
		similarly retained, but a larger residual contribution remains around the
		off-manifold component because of the broader Tikhonov tails.
		(d) Dispersion residual $E_{\mathrm{disp}}$ as a function of information
		retention $R_{\mathrm{info}}$ along the complete regularization paths.
		At the matched operating point $R_{\mathrm{info}}=0.573$, the Gaussian
		manifold projection gives $E_{\mathrm{disp}}=0.240$, compared with
		$E_{\mathrm{disp}}=0.268$ for Tikhonov regularization, corresponding to a
		$10.5\%$ lower residual at the same retained spectral information.}
	\label{fig6}
\end{figure}
\appendix
\onecolumngrid
\section*{End Matter}
\twocolumngrid
\textit{Numerical validation of the finite-width manifold projection}---For the numerical studies, the spectral variables are expressed in nondimensional form. Introducing characteristic scales $k_0$ and $\eta_0$, we define
\begin{equation}
	\hat{k}=\frac{k}{k_0},
	\qquad
	\hat{\eta}=\frac{\eta}{\eta_0},
	\qquad
	\hat{\alpha}=\frac{\alpha k_0^2}{\eta_0},
	\label{eq:nondimensionalization}
\end{equation}
such that the diffusion relation becomes
$
	\hat{\eta}=\hat{\alpha}\hat{k}^2.
$
The manifold width is correspondingly normalized as
$\hat{\sigma}=\sigma/\eta_0$.
Hats are omitted below for compactness; all reported values of
$\alpha$, $\eta$, and $\sigma$ are therefore dimensionless unless otherwise stated.
For comparison with conventional frequency-magnitude filtering, three spectral filters are considered:
\begin{align}
	W_{\mathrm{G}}(k)
	&=
	\exp\left[-\left(\frac{|k|}{k_c}\right)^2\right],
	\label{eq:gaussian_lowpass}
	\\
	W_{\mathrm{B}}(k)
	&=
	\frac{1}{
		1+\left(|k|/k_c\right)^{2n}},
	\label{eq:butterworth_lowpass}
\end{align}
together with the ideal low-pass truncation
\begin{equation}
	W_{\mathrm{T}}(k)
	=
	\begin{cases}
		1, & |k|\leq k_c,\\
		0, & |k|>k_c.
	\end{cases}
	\label{eq:hard_truncation}
\end{equation}
Unlike the proposed projection, these filters select spectral components according to $|k|$ and do not explicitly account for consistency with the diffusion relation.
The influence of the manifold width $\sigma$ is examined at a fixed input SNR of $10~\mathrm{dB}$. Figure~\ref{fig2} shows representative reconstructed spectra and the corresponding reconstruction metrics. A narrow projection width strongly suppresses off-manifold components but can also remove recoverable spectral information. Increasing $\sigma$ initially improves reconstruction fidelity by admitting the finite spectral neighborhood occupied by the physical signal. Beyond an intermediate range, however, the projection becomes increasingly permissive and admits more dispersion-inconsistent components. Accordingly, the PSNR exhibits a maximum and the relative $L_2$ error a minimum at an intermediate value of $\sigma$, whereas the dispersion residual $E_{\mathrm{disp}}$ increases as the constraint is relaxed.
To quantify the associated information loss, we define the information-retention ratio
$
	R_{\mathrm{info}}
	=
		\left\|S_{\mathrm{proj}}\right\|_2
	/
		\left\|S_{\mathrm{meas}}\right\|_2,
$
where $S_{\mathrm{meas}}$ and $S_{\mathrm{proj}}$ denote the measured and projected spectra, respectively. Small $R_{\mathrm{info}}$ indicates aggressive spectral rejection, whereas values approaching unity indicate that most of the measured spectral energy is retained.
Figure~\ref{fig2} make the resulting consistency--retention tradeoff explicit. Increasing $\sigma$ monotonically increases $R_{\mathrm{info}}$ but simultaneously increases $E_{\mathrm{disp}}$. Conversely, a very narrow manifold neighborhood minimizes the dispersion residual at the expense of substantial information loss. The relation between these quantities therefore defines a Pareto-type tradeoff between operator consistency and information retention.
The manifold width $\sigma$ can consequently be interpreted as a geometric regularization parameter. The optimum reconstruction is obtained in an intermediate regime in which the projection is sufficiently narrow to reject strongly dispersion-inconsistent components while remaining sufficiently broad to preserve the finite spectral thickness of the admissible signal. This behavior motivates the use of a finite-width soft projection rather than an exact zero-thickness manifold constraint.

The preceding analysis assumes that the diffusivity used to construct the manifold is known exactly. To examine sensitivity to material-parameter uncertainty, a reference spectrum is generated using the true diffusivity
$
	\alpha_0=0.10,
$
and corrupted by additive white Gaussian noise at an input SNR of $10~\mathrm{dB}$.
Projection is then performed using
$
	\alpha_{\mathrm{est}}=r\alpha_0,
	r=0.7,\;0.8,\;0.9,\;1.0,\;1.1,\;1.2,\;1.3,
$
corresponding to diffusivity errors between $-30\%$ and $+30\%$. The projection operator becomes
\begin{equation}
	W_{\sigma}
	\left(k,\eta;\alpha_{\mathrm{est}}\right)
	=
	\exp\left[
	-\frac{
		\left(\eta-\alpha_{\mathrm{est}}k^2\right)^2
	}{
		2\sigma^2
	}
	\right].
	\label{eq:mismatched_projection}
\end{equation}

Because the projection is now centered on an estimated rather than the true manifold, physical consistency is evaluated with respect to $\alpha_0$:
\begin{equation}
	E_{\mathrm{disp}}^{(\alpha_0)}(S)
	=
	\frac{
		\left\|
		\left(\eta-\alpha_0 k^2\right)S(k,\eta)
		\right\|_2
	}{
		\left\|S(k,\eta)\right\|_2
	}.
	\label{eq:true_dispersion_residual}
\end{equation}
The geometric displacement caused by parameter mismatch is additionally quantified as
\begin{equation}
	E_{\alpha}
	=
	\frac{
		\left\|
		\left(\alpha_{\mathrm{est}}-\alpha_0\right)
		k^2S_{\mathrm{true}}(k,\eta)
		\right\|_2
	}{
		\left\|S_{\mathrm{true}}(k,\eta)\right\|_2
	}.
	\label{eq:manifold_displacement}
\end{equation}
Thirty independent noise realizations are used for each diffusivity ratio, and the reported metrics are ensemble averages.
Figure~\ref{fig2} shows that the reconstruction error and true-manifold dispersion residual are minimized near
$\alpha_{\mathrm{est}}/\alpha_0=1$.
Underestimating or overestimating $\alpha$ displaces the projection manifold relative to the true spectral distribution and progressively attenuates physically admissible components. The degradation nevertheless remains moderate for small parameter errors, demonstrating that the finite-width projection provides tolerance to uncertainty in the assumed diffusivity.
This behavior admits a simple geometric interpretation. If the diffusivity lies within an uncertainty interval
$\alpha\in[\alpha_-,\alpha_+]$, the corresponding admissible manifolds form the set
\begin{equation}
	\mathcal{T}_{\alpha}
	=
	\bigcup_{\alpha\in[\alpha_-,\alpha_+]}
	\mathcal{M}_d(\alpha),
	\label{eq:manifold_tube}
\end{equation}
which constitutes a finite tube in spectral space. A soft projection can overlap this uncertainty-induced tube even when the assumed manifold is slightly displaced from the true one, whereas an exact zero-width projection would be substantially more sensitive to such mismatch.

A second source of effective spectral thickness arises from finite acquisition. Practical measurements are recorded over a finite temporal interval and at a finite sampling rate, limiting both temporal spectral resolution and the range of resolvable temporal scales. For an observation duration $T_{\mathrm{obs}}$, the characteristic Fourier-frequency resolution scales as
$
	\Delta\omega_{\mathrm{win}}
	\sim
	2\pi/T_{\mathrm{obs}},
$
while a sampling frequency $f_s$ imposes the Nyquist bound
$
	|\omega|\leq\pi f_s.
$
These Fourier-domain relations are not assumed to define a direct broadening law in the decay-rate coordinate $\eta$. Instead, finite acquisition is introduced numerically by truncating and discretely sampling the simulated temporal field before estimating its decay-rate representation.
Two projection strategies are compared. A nearly exact constraint is represented by
$
	\sigma_{\mathrm{exact}}=0.025,
$
whereas the proposed method employs a finite width $\sigma_{\mathrm{soft}}$. Both are applied at an input SNR of $20~\mathrm{dB}$.

Figure~\ref{fig2} demonstrates that shortening the observation interval broadens the estimated spectral distribution around the ideal diffusion relation. A nearly exact projection strongly concentrates the reconstructed spectrum around the theoretical manifold and therefore produces a small dispersion residual, but it simultaneously rejects a substantial fraction of the broadened signal. The finite-width projection preserves a larger portion of this acquisition-broadened spectral content and consequently yields lower reconstruction error.
The same behavior persists as the sampling rate is reduced. Although reconstruction quality deteriorates for both strategies, the finite-width projection remains less sensitive to acquisition-induced spectral distortion.
Finite observation and sampling therefore provide an acquisition-based reason for avoiding an exact zero-thickness manifold constraint. Even when the underlying field obeys the diffusion equation, its finite-resolution spectral representation need not lie exactly on the ideal operator manifold. The physically useful admissible set is consequently better represented as a finite neighborhood whose width accounts for both intrinsic spectral spread and measurement-induced uncertainty.
\begin{table*}[t]
	\centering
	\caption{
		Experimental comparison of conventional spatial low-pass filtering and
		dispersion-constrained manifold projection. The Gaussian, Butterworth, and
		hard-cutoff filters were adjusted to match the field-retention level of the
		representative manifold projection, $R_{\mathrm{field}}=0.952$ for
		$\sigma=8\Delta\eta$. The SNR gain is calculated relative to the unprocessed
		measurement. The last column reports the SNR evaluated at the peak time of
		the original measurement, $t=0.30~\mathrm{s}$, thereby providing a
		fixed-time comparison independent of peak-time shifts.
	}
	\label{tab:supp_exp_snr}
	\begin{ruledtabular}
		\begin{tabular}{lcccccc}
			Method
			& $R_{\mathrm{field}}$
			& $k_c$ (rad m$^{-1}$)
			& Peak SNR (dB)
			& Gain (dB)
			& $t_{\mathrm{peak}}$ (s)
			& SNR at $0.30$ s (dB)
			\\
			\hline
			Original
			& 1.000
			& --
			& 9.65
			& 0.00
			& 0.30
			& 9.65
			\\
			
			Gaussian LPF
			& 0.952
			& $2.001\times10^{4}$
			& 16.83
			& 7.18
			& 0.30
			& 16.83
			\\
			
			Butterworth LPF
			& 0.952
			& $1.500\times10^{4}$
			& 16.88
			& 7.24
			& 0.30
			& 16.88
			\\
			
			Hard cutoff
			& 0.952
			& $1.401\times10^{4}$
			& 16.90
			& 7.25
			& 0.30
			& 16.90
			\\
			
			Manifold projection
			& 0.952
			& --
			& \textbf{23.13}
			& \textbf{13.48}
			& 0.42
			& \textbf{21.95}
			\\
		\end{tabular}
	\end{ruledtabular}
\end{table*}

\textit{Comparison with operator-residual regularization}---The low-pass-filter comparison in the main text is designed to demonstrate that spectral magnitude and operator consistency constitute distinct selection criteria, rather than to establish quantitative superiority over all regularization strategies. To provide a more stringent physics-informed comparison, we additionally consider a spectral Tikhonov regularization constructed from the same diffusion-operator residual
$
	r_d(\mathbf{k},\eta)
	=
	\eta-\alpha |\mathbf{k}|^2.
$
Specifically, we define
\begin{equation}
	S_{\mathrm{Tik}}
	=
	\underset{S}{\operatorname{arg\,min}}
	\left[
	\left\|S-S_{\mathrm{meas}}\right\|_2^2
	+
	\lambda
	\left\|r_d S\right\|_2^2
	\right],
	\label{eq:tikhonov}
\end{equation}
where $\lambda$ controls the strength of the operator-residual penalty. Because the optimization is diagonal in $(\mathbf{k},\eta)$ space, its solution is
$
	S_{\mathrm{Tik}}(\mathbf{k},\eta)
	=
	W_{\mathrm{Tik}}(\mathbf{k},\eta)
	S_{\mathrm{meas}}(\mathbf{k},\eta),
	W_{\mathrm{Tik}}(\mathbf{k},\eta)
	=
	1/
	(1+\lambda r_d^2(\mathbf{k},\eta)).
$
This baseline therefore incorporates exactly the same diffusion relation as the proposed finite-width projection,
$
	W_{\sigma}(\mathbf{k},\eta)
	=
	\exp\left[
	-r_d^2(\mathbf{k},\eta)
	/(2\sigma^2)
	\right],
$
but imposes a conventional quadratic residual penalty rather than a Gaussian finite-width neighborhood around the diffusion manifold. The comparison consequently isolates the effect of the spectral regularization geometry from the use of diffusion physics itself.
To ensure a fair comparison, $\lambda$ is selected such that the Tikhonov reconstruction has approximately the same information-retention ratio
$
	R_{\mathrm{info}}
	=
	\left\|S_{\mathrm{rec}}\right\|_2/
	\left\|S_{\mathrm{meas}}\right\|_2
$
as the corresponding manifold-projected spectrum. Reconstruction quality is then evaluated using PSNR, relative $L_2$ error, and the dispersion residual $E_{\mathrm{disp}}$. This matched-retention comparison avoids attributing an apparent improvement simply to stronger spectral attenuation and directly tests whether the finite-width manifold weighting provides a different reconstruction behavior from conventional operator-residual regularization.
Figures~\ref{fig6}(b) and
\ref{fig6}(c) show representative results at
$R_{\mathrm{info}}=0.573$. Both methods preserve the high-wavenumber
component located on the reference diffusion manifold. The difference is
most apparent for the deliberately introduced off-manifold component:
because of its broader residual tails, the Tikhonov weighting retains a
larger fraction of this component, whereas the Gaussian finite-width
weighting suppresses it more strongly.
This behavior is quantified in Fig.~\ref{fig6}(d), which
plots the dispersion residual against information retention along the full
regularization paths. At the matched operating point
$R_{\mathrm{info}}=0.573$, the Gaussian manifold projection yields
$E_{\mathrm{disp}}=0.240$, whereas operator-residual Tikhonov
regularization gives $E_{\mathrm{disp}}=0.268$. Thus, for the same retained
spectral information, the finite-width Gaussian weighting produces a
$10.5\%$ lower operator residual.

\textit{Experimental comparison with conventional spatial filtering}---To determine whether the experimentally observed SNR enhancement can be
attributed simply to generic spatial smoothing, we further compare the
dispersion-constrained manifold projection with three conventional spatial
low-pass filters: a Gaussian filter, a Butterworth filter, and an ideal
hard spectral cutoff. Importantly, the comparison is performed at matched
field retention rather than by independently optimizing each filter for
maximum SNR. This prevents the comparison from favoring a method merely
because it removes a larger fraction of the measured field.
The experimental signal-to-noise ratio is evaluated directly in the
temperature field and therefore does not involve the operator residual
$r_d$ used to construct the manifold weighting. For each time frame, a
signal region $\Omega_s$ containing the defect and a defect-free background
region $\Omega_b$ are defined. The corresponding mean temperatures are
\begin{equation}
	\mu_s(t)
	=
	\frac{1}{|\Omega_s|}
	\sum_{\mathbf{x}\in\Omega_s}
	T(\mathbf{x},t),
\end{equation}
and
\begin{equation}
	\mu_b(t)
	=
	\frac{1}{|\Omega_b|}
	\sum_{\mathbf{x}\in\Omega_b}
	T(\mathbf{x},t).
\end{equation}

The background fluctuation is quantified by the sample standard deviation
\begin{equation}
	\sigma_b(t)
	=
	\left[
	\frac{1}{|\Omega_b|-1}
	\sum_{\mathbf{x}\in\Omega_b}
	\left(
	T(\mathbf{x},t)-\mu_b(t)
	\right)^2
	\right]^{1/2}.
\end{equation}

The frame-dependent thermal SNR is then defined as Eq~\eqref{eq:thermal_snr}.
Two quantities are reported below. The first is the peak SNR,
$
	\mathrm{SNR}_{\mathrm{peak}}
	=
	\max_t \mathrm{SNR}(t),
$
which allows each processing method to reach its own optimal observation
time. The second is the SNR evaluated at the peak time of the original
measurement, $t=0.30~\mathrm{s}$. The latter provides a stricter fixed-time
comparison and eliminates any advantage associated with a processing-induced
shift of the SNR maximum.

Three conventional filterings were introduced in Eqs~\eqref{eq:butterworth_lowpass}--~\eqref{eq:hard_truncation}.
A direct comparison at arbitrarily selected cutoff frequencies would not be
well controlled because a stronger low-pass filter can trivially reduce
background fluctuations by removing more spectral content. We therefore
match all conventional filters to the field-retention level of the
representative manifold projection used in the experimental analysis,
$\sigma=8\Delta\eta$.
For a processed temperature field $T_f$, we define the field retention as
$
	R_{\mathrm{field}}
	=
		\|T_f\|_2
	/
		\|T_{\mathrm{raw}}\|_2,
$
where the norm is evaluated over all spatial pixels and all measured time
frames. The representative manifold projection gives
$
	R_{\mathrm{field}}^{\mathrm{manifold}}
	=
	0.952173.
$
The cutoff $k_c$ of each conventional filter is therefore determined
numerically from
$
	R_{\mathrm{field}}^{(f)}
	=
	R_{\mathrm{field}}^{\mathrm{manifold}},
$
rather than from its resulting SNR.
For computational efficiency, the matching is performed directly in the
spatial Fourier domain using Parseval's theorem. Defining the accumulated
spatial spectral energy
$
	P(\mathbf{k})
	=
	\sum_t
	\left|
	\widetilde{T}(\mathbf{k},t)
	\right|^2,
$
the retention associated with a spatial weighting $W_f$ can be evaluated as
\begin{equation}
	R_{\mathrm{field}}^{(f)}
	=
	\left[
	\frac{
		\sum_{\mathbf{k}}
		|W_f(\mathbf{k})|^2
		P(\mathbf{k})
	}{
		\sum_{\mathbf{k}}
		P(\mathbf{k})
	}
	\right]^{1/2}.
	\label{eq:supp_parseval_retention}
\end{equation}

The cutoff is then varied until 
$
R_{\mathrm{field}}^{(f)}
=
R_{\mathrm{field}}^{\mathrm{manifold}},
$ is satisfied.
This procedure gives
$
	k_c^{\mathrm{G}}
	=
	2.001\times10^4~\mathrm{rad\,m^{-1}},
	k_c^{\mathrm{B}}
	=
	1.500\times10^4~\mathrm{rad\,m^{-1}},
	k_c^{\mathrm{H}}
	=
	1.401\times10^4~\mathrm{rad\,m^{-1}},
$
for the Gaussian, sixth-order Butterworth, and hard-cutoff filters,
respectively. The resulting retention is $R_{\mathrm{field}}=0.952173$
for all three filters to numerical precision. No SNR information is used
in selecting these cutoff values.

Table~\ref{tab:supp_exp_snr} summarizes the resulting experimental SNR
performance. The unprocessed measurement exhibits a peak SNR of
$9.65~\mathrm{dB}$ at $t=0.30~\mathrm{s}$. All three conventional
low-pass filters improve the peak SNR, confirming that removal of
high-spatial-frequency fluctuations provides a substantial generic
denoising benefit. At the common field-retention level
$R_{\mathrm{field}}=0.952$, the Gaussian, Butterworth, and hard-cutoff
filters yield peak SNR values of $16.83$, $16.88$, and $16.90~\mathrm{dB}$,
respectively, corresponding to gains of approximately $7.2~\mathrm{dB}$
relative to the original measurement.
In contrast, the dispersion-constrained manifold projection with
$\sigma=8\Delta\eta$ reaches a peak SNR of $23.13~\mathrm{dB}$, giving an
improvement of $13.48~\mathrm{dB}$ relative to the unprocessed field.
Thus, despite retaining essentially the same total field norm as the
conventional filters, the manifold projection provides approximately
$
	6.23~\mathrm{dB}
$
higher peak SNR than the best-performing conventional spatial filter in
this comparison.
The peak of the manifold-processed SNR occurs at $t=0.42~\mathrm{s}$,
whereas the maxima of the unprocessed and conventional low-pass-filtered
fields occur at $t=0.30~\mathrm{s}$. To verify that the higher peak SNR is
not merely caused by this shift in optimal observation time, we additionally
compare all methods at the fixed time $t=0.30~\mathrm{s}$. At this common
time, the manifold projection gives an SNR of $21.95~\mathrm{dB}$, compared
with $16.83$--$16.90~\mathrm{dB}$ for the three conventional filters.
Therefore, even without allowing the manifold projection to select its own
optimal time, it retains an SNR advantage of approximately $5.1~\mathrm{dB}$.
The close agreement among the three conventional filters is also
informative. Although their transfer functions differ substantially in
shape---from smooth Gaussian attenuation to a sixth-order Butterworth
transition and an abrupt hard cutoff---their matched-retention SNR values
differ by less than $0.1~\mathrm{dB}$. This indicates that, for the present
experiment, the principal limitation is not the detailed shape of the
low-pass transition. Rather, all three methods impose essentially the same
ordering principle: large-$k$ components are preferentially attenuated
irrespective of their consistency with the reference diffusion operator.
The manifold projection uses a fundamentally different selection
coordinate. Instead of weighting a component according to $k$ alone, its
spectral weighting is constructed from the residual
$
	r_d(\mathbf{k},\eta)
	=
	\eta-\alpha|\mathbf{k}|^2,
$
through Eq~\eqref{eq:weight}.
Consequently, a large-$k$ component is not suppressed merely because it
corresponds to a fine spatial scale; it can be retained when its decay rate
is consistent with the reference diffusion relation. Conversely, a
lower-$k$ component can be attenuated when its measured decay-rate content
lies far from the reference manifold. The conventional filters in
Eqs.~\eqref{eq:gaussian_lowpass}--\eqref{eq:hard_truncation} cannot make
this distinction because their weights contain no decay-rate coordinate.
The matched-retention experiment therefore separates two possible sources
of SNR improvement. Approximately $7.2~\mathrm{dB}$ of improvement can be
obtained through conventional spatial low-pass filtering at the prescribed
retention level, demonstrating the expected benefit of generic smoothing.
The additional improvement obtained by the manifold projection cannot,
however, be explained solely by stronger attenuation, because the compared
fields retain the same global $L_2$ norm. Instead, the result is consistent
with selective redistribution of the retained spectral content according
to reference-operator consistency rather than spatial scale alone.

\end{document}